\documentclass[sigconf]{acmart}

\usepackage{makecell}
\usepackage{algorithmic}
\usepackage{textcomp}
\usepackage{subfigure}
\usepackage{xspace}
\usepackage{mathtools}
\usepackage{enumitem}
\usepackage{multirow}
\usepackage{siunitx}
\usepackage{verbatim}
\usepackage{listings}

\copyrightyear{2026}
\acmYear{2026}
\setcopyright{cc}
\setcctype{by-nc-nd}
\acmConference[MSR '26]{23rd International Conference on Mining Software Repositories}{April 13--14, 2026}{Rio de Janeiro, Brazil}
\acmBooktitle{23rd International Conference on Mining Software Repositories (MSR '26), April 13--14, 2026, Rio de Janeiro, Brazil}
\acmPrice{}
\acmDOI{10.1145/3793302.3793577}
\acmISBN{979-8-4007-2474-9/2026/04}

\begin{document}
\title{How do Agents Refactor: An Empirical Study}

\settopmatter{authorsperrow=4}

\author{Lukas Ottenhof}
\affiliation{%
  \institution{University of Alberta}
  \city{Edmonton}
  \state{AB}
  \country{Canada}
}
\email{lottenho@ualberta.ca}

\author{Daniel Penner}
\affiliation{%
  \institution{University of Alberta}
  \city{Edmonton}
  \state{AB}
  \country{Canada}
}
\email{dpenner2@ualberta.ca}

\author{Abram Hindle}
\affiliation{%
  \institution{University of Alberta}
  \city{Edmonton}
  \state{AB}
  \country{Canada}
}
\email{hindle1@ualberta.ca}

\author{Thibaud Lutellier}
\affiliation{%
  \institution{University of Alberta}
  \city{Edmonton}
  \state{AB}
  \country{Canada}
}
\email{lutellie@ualberta.ca}

\begin{abstract}
\label{sec:abstract}

Software development agents such as Claude Code, GitHub Copilot, Cursor Agent, Devin, and OpenAI Codex are being increasingly integrated into developer workflows. While prior work has evaluated agent capabilities for code completion and task automation, there is little work investigating how these agents perform Java refactoring in practice, the types of changes they make, and their impact on code quality. In this study, we present the first analysis of agentic refactoring pull requests in Java, comparing them to developer refactorings across 86 projects per group. Using RefactoringMiner and DesigniteJava 3.0, we identify refactoring types and detect code smells before and after refactoring commits. Our results show that agent refactorings are dominated by annotation changes (the 5 most common refactoring types done by agents are annotation related), in contrast to the diverse structural improvements typical of developers. Despite these differences in refactoring types, we find Cursor to be the only model to show a statistically significant increase in refactoring smells.

\end{abstract}



\maketitle

\section{Introduction}
\label{sec:intro}
Agents are becoming an increasingly important part of modern software development. With the rise of code agents such as Claude Code~\cite{anthropic2025claudecode}, Copilot~\cite{microsoft2025copilot}, Cursor Agent~\cite{cursor2025docs}, Devin~\cite{devin2025ai} and OpenAI Codex~\cite{openai2025codex}, developers are increasingly using agents throughout the development process. Recent work shows agents are increasingly used on GitHub. Within just a few months of their introduction, the combined number of commits and pull requests made by Claude Code, Cursor Agent, and Codex already exceeded 1.3 million~\cite{zi2025agentpack}. 

Studies show developers using agents such as GitHub Copilot can complete tasks significantly faster than those without agent assistance~\cite{peng2023the}. Prior work has attempted to apply agents and examined their ability to complete tasks such as programming ~\cite{peng2023the,wang2025ai,humaninloop}. However, work investigating how agents perform Java refactoring, the quality of their refactorings, and the types of refactoring they carry out in practice remains limited.

Refactoring is an important aspect of software maintenance, aimed at improving code quality, readability, and technical debt~\cite{refactoring1, fowler2018refactoring}. However, refactoring is often perceived as risky, potentially introducing regression bugs or build breaks~\cite{kim2012field}, and can lead to increased code smells. Smells indicate deeper design or implementation problems in code quality~\cite{palomba2018diffuseness}. Considering these perceived risks, understanding how agents perform refactoring, which refactoring types they introduce, and if their refactoring commits are improving code quality can show both their current capabilities and their limitations when contributing to software repositories. Additionally, knowing what types of refactorings agents are performing in practice helps developers to know what to expect when integrating agents into their workflows.

In this paper, we present an analysis of agentic refactoring pull requests on GitHub. Our goal is to understand how five commonly used agents (Claude Code~\cite{anthropic2025claudecode}, Copilot~\cite{microsoft2025copilot}, Cursor Agent~\cite{cursor2025docs}, Devin~\cite{devin2025ai}, and OpenAI Codex~\cite{openai2025codex}) are modifying code, the types of refactorings they perform, and whether or not their pull requests containing refactorings improve code quality. By examining both the frequency and nature of code refactorings, we provide insights into the current refactoring capabilities and limitations of agents.

During our evaluation, we aimed to answer two research questions: \textbf{RQ1:} What types of refactoring are agents doing, how frequent is agent refactoring? \textbf{RQ2:} Are agents and humans equally likely to introduce code smells when refactoring?

\section{Data Processing}
\label{sec:approach}
We extracted two Java pull request datasets from AIDev~\cite{aiDev} (accessed October 20th 2025): an agentic dataset and a developer dataset. Java was chosen because it is widely analyzed in software engineering research and well supported by static analysis tools for refactoring and code smell detection~\cite{codeSmells, refactoringMiner, refactoringMiner1, refactoringMiner2}. Using RefactoringMiner, we identified refactoring commits within pull requests and analyzed code smells before and after each refactoring commit to assess refactoring types and quality. Although refactorings could occur in any file, code smells were measured only in ``.java'' files. Table \ref{tab:commit_refactoring_per_age} lists the total and refactoring commits in each dataset (commits with at least one refactoring type).

\subsection{Agentic Dataset}
To create an agentic Java pull request dataset, we used the AIDev dataset~\cite{aiDev}, which contains 932,791 agentic pull requests from five code agents: Claude Code, Cursor, Devin, GitHub Copilot, and OpenAI Codex. From this dataset, we extracted all 1,278 pull requests made by agents in the 86 Java projects present to analyze how agents perform Java refactoring in practice. From these pull requests, we extracted 2,626 total agentic commits. To detect refactoring instances, we used RefactoringMiner ~\cite{refactoringMiner, refactoringMiner1, refactoringMiner2}, a state of the art refactoring detection tool with an overall precision of 97.96\%, recall of 87.20\%, and F1 score of 92.27\% across refactoring types. Using refactoring miner on the agentic commits, we found 413 commits had refactoring changes (324 refactoring commits where ``.java'' files were changed). To analyze code smells introduced by refactoring, we used DesigniteJava 3.0 ~\cite{codeSmells} which has a precision, recall, and F1 score of 98.15\%, 99.5\%, and 98.82\%, respectively. 

\subsection{Developer Dataset}  
To provide a baseline for comparison with agentic refactoring, we collected a dataset of 86 Java projects from GitHub which do not contain agent activity. Each project in the baseline has at least 50 stars and has not been updated since January 2021, ensuring minimal agentic or AI-assisted influence on the code. Of these 86 projects, 82 had pull requests (6,466 commits, 958 refactoring commits, and 2,537 pull requests). This baseline allows us to give context to the quality and frequency of agent refactoring relative to traditional, developer-generated ones. Similar to the agentic dataset, to detect code refactoring we used RefactoringMiner~\cite{refactoringMiner, refactoringMiner1, refactoringMiner2}, and to detect code smells we used DesigniteJava 3.0 ~\cite{codeSmells}.

\begin{table}[h]
\setlength{\tabcolsep}{4pt}
\centering
\small
\begin{tabular}{l r r r c}
\toprule
\textbf{Agent} & \textbf{\# Commits} & \textbf{\# Ref. Commits} & \textbf{\% Rate} & \textbf{95\% Rate CI} \\ 
\midrule
Claude  & 37   & 22   & 59.46 & [43.24, 75.68] \\ 
Copilot & 1190 & 156  & 13.11 & [11.18, 15.13] \\ 
Cursor  & 68   & 7    & 10.29 & [4.11, 17.65]  \\ 
Devin   & 41   & 5    & 12.20 & [2.44, 21.95]  \\ 
OpenAI  & 1290 & 223  & 17.29 & [15.27, 19.46] \\ 
Developers   & 6466 & 958 & 14.82 & [13.90, 15.70] \\ 
\bottomrule
\end{tabular}
\caption{Total and refactoring commits per agent. 95\% confidence intervals (CI) for refactoring rates were calculated using bootstrapping with replacement (2,000 iterations).}
\Description{A table showing the number of commits, number of refactoring commits, percentage rate, and 95 percent confidence interval for six agents: Claude, Copilot, Cursor, Devin, OpenAI, and human developers. Claude has the highest refactoring rate, while Cursor and Devin have the lowest.}
\label{tab:commit_refactoring_per_age}
\end{table}
\section{Evaluation and Results}
\label{sec:results}

\subsection{RQ1: What types of refactoring are agents doing, how frequent is agent refactoring?}
We analyzed the distribution and frequency of refactoring types across our agentic and developer datasets to analyze the difference in refactoring done by agents and developers in practice. Overall, we found clear differences between developer and agentic refactoring: agent refactoring is often dominated by a few refactoring types, and developers have a very diverse set of refactoring they do. Furthermore, within the agentic dataset, implemented refactoring types varied greatly between agents.

In our developer dataset, refactorings were diverse; no single type dominated the refactoring type distribution. The most common refactoring, Move Class, accounted for only 6.38\% of all refactorings, followed by Change Attribute Access Modifier (6.26\%) and Extract Method (5.09\%). This distribution shows developers have a tendency to perform a variety of structural improvements to enhance design and maintainability, rather than focusing on a single refactoring pattern, such as annotation-related changes. 

Refactoring types varied greatly across agents. For instance, over 91\% of Claude Code’s refactorings were annotation related. In contrast to developers, whose refactorings are diverse with no dominant type (most frequent is 6.38\%), many agents concentrate on a few annotation related refactorings. Overall, Add Method Annotation (22.52\%), Add Parameter Annotation (12.82\%), and Modify Method Annotation (10.37\%) dominate agent refactorings, with a sharp drop to the next highest at 4.42\%. Unlike developers, all agents had annotation changes in their top five most common refactorings, suggesting agent refactorings are heavily concentrated on annotation edits rather than structural changes. Table \ref{tab:humans-vs-agents-top} shows the 10 most common refactoring actions done by agents and developers. Additionally, agents performed significantly more refactoring actions in refactoring commits than developers (see Table \ref{tab:commit_refactoring_rate}).
\begin{table}[h]
\resizebox{\columnwidth}{!}{
\begin{tabular}{c l c c}
\toprule
\textbf{Rank} & \textbf{Refactoring Type} & \textbf{Count} & \textbf{Share (\%)} \\
\midrule
\multicolumn{4}{l}{\textbf{Developers — Top 10 by Share}} \\
1  & Move Class & 744 & 6.38 \\
2  & Change Attribute Access Modifier & 730 & 6.26 \\
3  & Extract Method & 594 & 5.09 \\
4  & Rename Method & 537 & 4.61 \\
5  & Change Parameter Type & 511 & 4.38 \\
6  & Change Method Access Modifier & 507 & 4.35 \\
7  & Change Return Type & 461 & 3.95 \\
8  & Add Method Annotation & 419 & 3.59 \\
9  & Change Attribute Type & 402 & 3.45 \\
10 & Add Parameter & 398 & 3.41 \\
\midrule
\multicolumn{4}{l}{\textbf{Agentic — Top 10 by Share}} \\
1  & Add Method Annotation & 8,150 & 22.52 \\
2  & Add Parameter Annotation & 4,639 & 12.82 \\
3  & Modify Method Annotation & 3,753 & 10.37 \\
4  & Remove Method Annotation & 1,600 & 4.42 \\
5  & Modify Class Annotation & 1,544 & 4.27 \\
6  & Change Variable Type & 1,511 & 4.17 \\
7  & Remove Parameter & 1,453 & 4.01 \\
8  & Move Class & 1,377 & 3.80 \\
9  & Rename Variable & 981 & 2.71 \\
10 & Rename Parameter & 838 & 2.32 \\
\bottomrule
\end{tabular}
}
\caption{Top 10 refactoring types by count and share for
Developers (N = 11,659) and Agentic models (N = 36,196).}
\Description{A table listing the top 10 refactoring types by count and share for developers and agentic models. For developers, the most frequent refactorings include changing attribute access modifiers, moving classes, and extracting methods, each comprising 3 to 6 percent of total refactorings. For agents, the most common refactorings are adding method annotations, adding parameter annotations, and modifying method annotations. The table compares absolute counts and percentage share of each refactoring type.}
\label{tab:humans-vs-agents-top}
\end{table}

Claude Code had the most skewed refactoring distribution among all agents. Out of 16,780 total refactorings, over 91\% were annotation-related, with Add Method Annotation (41.9\%), Add Parameter Annotation (27.6\%), and Modify Method Annotation (21.4\%) dominating its changes. This indicates that Claude’s refactoring behaviour is highly consistent, primarily editing annotations rather than altering functionality or structure. Claude also performed a high number of changes per commit with a mean of 762.73 and a median of 475, suggesting it is doing large batches of annotation updates.

Copilot had more diversity in the types of refactoring changes it made compared to other agents. Its top refactorings include Remove Parameter (9.1\%), Move Class (8.7\%), Remove Method Annotation (7.9\%), and Modify Class Annotation (7.5\%). These indicate a mix of structural edits and annotation adjustments, aligning more closely with developer refactoring. 

Cursor Agent performed the fewest refactorings overall (31), despite being used in 9 projects. Its focus was on structural changes. Cursor's most common refactoring type was Extract Method (29.0\%), followed by Extract and Move Method (16.13\%) and Parameterize Variable (16.13\%). These indicate that Cursor is used to implement code restructuring as opposed to annotation related changes.
    \begin{table}[h]
    \setlength{\tabcolsep}{4pt} 
    \centering
    \small
    \begin{tabular}{l r c r r r r}
    \toprule
    \textbf{Agent} & \textbf{Mean} & \textbf{95\% Mean CI} & \textbf{Med} & \textbf{Std} & \textbf{Min} & \textbf{Max} \\
    \midrule
    Claude  & 762.73 & [425.55, 1099.17] & 475 & 830.90 & 1 & 2320 \\ 
    Copilot & 100.74 & [52.22, 156.00]   & 4   & 337.75 & 1 & 2646 \\ 
    Cursor  & 4.43   & [2.28, 6.71]      & 4   & 3.26   & 1 & 10   \\ 
    Devin   & 14.00  & [7.59, 20.00]     & 14  & 7.87   & 2 & 24   \\ 
    OpenAI  & 16.14  & [8.88, 26.22]     & 2   & 65.20  & 1 & 771  \\ 
    Developers& 12.17 &[10.49, 14.03] & 3   & 28.85  & 1 & 428  \\ 
    \bottomrule
    \end{tabular}
    \caption{Refactors per refactoring commit by agent (2,000-iteration bootstrap CI with replacement).}
    \Description{Table showing the number of refactors per refactoring commit for each agent and developers. Claude shows the highest mean number of refactors per commit, while Cursor and Devin have much lower counts.}
    \label{tab:commit_refactoring_rate}
    \end{table}

Devin's refactorings were primarily related to annotations. The distribution was dominated by Add Attribute Annotation (32.9\%) and Remove Attribute Annotation (32.9\%), which are types of refactorings consistent with Claude Code. This suggests that like Claude Code, when making code changes, Devin is often making changes related to annotations rather than functionality.

Similar to Copilot, OpenAI Codex displayed diversity in refactoring types more in line with our developer baseline compared to other agents, with Rename Parameter (15.4\%) and Change Variable Type (9.5\%) as its most frequent refactorings. However, unlike Copilot and our developer baseline, Codex is performing large amounts of diverse annotation related refactoring when refactoring code.

\textbf{RQ1 Summary:} Agentic refactoring is dominated by annotation changes, especially for Claude Code. In contrast, developers exhibit a more balanced and diverse refactoring pattern, with a heavier focus on structural improvements such as class movement and method extraction.

\subsection{RQ2: Are agents and humans equally likely to introduce code smells when refactoring?}

To assess whether agents introduce or remove code smells when refactoring, we compared the number of detected code smells before and after each refactoring commit using DesigniteJava 3.0~\cite{codeSmells}. Table~\ref{tab:smell-before-after-with-delta} summarizes the overall smell counts, and the mean change in counts. A positive $\Delta$ indicates the refactoring introduced additional smells, and a negative value indicates smell removal. Across all data       sets, most agents exhibited small to moderate increases in code smells following refactoring, indicating that while agents perform large amounts of refactoring, they do not consistently improve overall code quality. The most common smells introduced by agents were Assertion roulette (15\%), Long Statement (8\%), and Unnecessary Abstraction (8\%), while the most common smells introduced by developers were Long Statement (8\%), Deficient Encapsulation (8\%), and Magic Number (7\%).

In the developer dataset, refactoring commits increased the mean number of code smells from 44.2 to 46.6 per commit, resulting in an average $\Delta$ of 2.43 (see Table~\ref{tab:smell-before-after-with-delta}). This minimal increase, with a median $\Delta$ of 0, suggests developer refactorings have little impact on code smells. While developers occasionally introduce new smells (maximum $\Delta$ of 199), these cases are outliers, and most developer refactorings maintain code health. This shows that although developers are generally making diverse structural changes in refactoring commits, these changes are not impacting code smells.

Claude Code had the largest average increase in code smells among all agents, rising from 179.1 to 214.6 per commit (mean $\Delta = 35.5$). However, the Wilcoxon rank sum test indicates this increase is not statistically significant compared to developers (p = 0.25), and the effect size is small (Cliff's Delta = 0.14). This suggests that while Claude Code performs large-scale annotation changes, their impact on code smells is consistent with developer refactorings.

GitHub Copilot showed minor changes in smell counts, increasing from 71.82 to 74.04 per commit (mean $\Delta = 2.22$). The difference from developers is not statistically significant (p = 0.42) with a negligible effect size (Cliff's Delta = -0.04), indicating Copilot’s refactorings generally maintain code quality similar to developers.

Cursor’s refactorings introduced a moderate increase in code smells, with a mean increase of 87.14 to 107.0 per commit (mean $\Delta = 19.86$, median $\Delta = 3.0$). Unlike other agents, this increase is statistically significant (p = 0.013) with a large effect size (Cliff's Delta = 0.51), suggesting that Cursor tends to introduce significantly more smells per refactoring commit than developers.

Devin did not show a significant change in code smells before and after refactoring (mean $\Delta = 0.0$), and the difference from developers is not significant (p = 0.36, Cliff's Delta = -0.22). Due to the very small sample size, no reliable conclusions can be drawn.

OpenAI Codex showed a minimal reduction in code smells on average, from 470.45 to 469.26 per commit (mean $\Delta = -1.2$). The difference compared to developers is not statistically significant (p = 0.56) with a negligible effect size (Cliff's Delta = -0.03), indicating Codex’s refactorings largely maintain code quality.

\begin{table}[h]
\small
\setlength{\tabcolsep}{4pt}
\centering

\begin{tabular}{p{1.3cm} p{1.05cm} p{1.05cm} p{1.05cm} p{1.05cm} p{1.05cm}}
\toprule
\textbf{Agent} &
\textbf{Mean$_B$} & \textbf{Med$_B$} &
\textbf{Mean$_A$} & \textbf{Med$_A$} &
\textbf{Mean$_\Delta$} \\
\midrule
Claude & 179.10 & 126 & 214.60 & 211 & 35.50 \\
Copilot & 71.82 & 26 & 74.04 & 28 & 2.22 \\
Cursor & 87.14 & 3 & 107.00 & 21 & 19.86 \\
Devin & 60.00 & 62 & 60.00 & 62 & 0.00 \\
OpenAI & 470.45 & 29 & 469.26 & 31 & -1.20 \\
Developers & 44.20 & 15 & 46.63 & 16 & 2.43 \\
\bottomrule
\end{tabular}%

\caption{Summary of detected smells before (B), and after (A) refactoring commit, and the mean change across agents.}
\Description{Table showing the mean and median number of code smells detected before and after refactoring commits for each agent and developers. Claude has the largest increase in smells, while Devin and OpenAI show minimal changes. Copilot and Developers have small increases.}
\label{tab:smell-before-after-with-delta}
\end{table}

To understand differences in agent performance regarding the impact of refactoring commits on code smells, we used the Wilcoxon rank sum test along with Cliff's Delta to compare each agent to our developer baseline. We ran these tests on the code smells delta of each agent's refactoring commits. Across most agents, differences were statistically insignificant, despite some large mean differences (e.g., Claude Code, mean $\Delta = 35.5$ vs.\ developer mean $\Delta = 2.43$; Copilot, mean $\Delta = 2.22$; Devin, mean $\Delta = 0$; OpenAI Codex, mean $\Delta = -1.2$). This is reflected in the Wilcoxon rank-sum p-values (Claude Code = 0.25, Copilot = 0.42, Devin = 0.36, OpenAI Codex = 0.56) and the relatively small Cliff's Delta values (0.14, -0.04, -0.22, -0.03, respectively), indicating weak or negligible effect sizes. Cursor is the only agent with a statistically significant difference in code smell changes (mean $\Delta = 19.86$ vs.\ developer 2.43, p = 0.013, Cliff's Delta = 0.51). The combination of statistical significance and a moderate Cliff's Delta implies that refactoring commits by this agent generally decrease code quality. Conversely, the remainder of these results indicate that although agents may perform different types of refactorings, the overall distribution of code smell changes per commit is largely consistent with the developer baseline.

\textbf{RQ2 Summary:}
Although refactoring is typically used to improve code quality, developer refactoring commits slightly increase code smells on average. Most agents produce changes in code smells that are statistically indistinguishable from developers, despite some large mean differences in refactoring types. Cursor shows a statistically significant increase in code smells, while OpenAI Codex is the only agent to exhibit a decrease. Overall, these results indicate that although agents perform different types of refactorings, their impact on code smells per commit is consistent with developers.

\section{Implications}
\label{sec:implications}

While developers have a balanced distribution of refactorings, agents like Claude Code show a skew, with over 91\% of changes being annotation-related. This difference indicates that although agents perform more refactorings than humans, these changes do not necessarily have the same structural impact as human refactorings. The number of refactorings in an agent pull request should therefore not be treated as evidence of meaningful design improvement. In practice, maintainers should review agent refactoring pull requests by explicitly checking for structural changes (e.g., method extraction or class movement) rather than assuming large refactoring volumes imply architectural improvement. Moreover, because agent refactorings do not consistently reduce code smells and in some cases increase them, agent generated refactoring commits should be subjected to the same quality checks as human ones.

\section{Threats to Validity}
\label{sec:limits}

Our analysis uses agentic data from the AIDev dataset~\cite{aiDev}, which contains 932,791 agentic pull requests. However, Claude Code, Cursor, Devin, and OpenAI Codex appear in fewer than 10 Java repositories each, limiting generalizability for them. To mitigate this limited and biased representation of some agents, we report results for each agent individually rather than aggregating them.

Due to the dataset size, manual labeling is impractical, so we rely on automated tools for refactoring and code smell detection. Consequently, our results depend on tool accuracy. To mitigate this threat, we use state of the art tools: RefactoringMiner~\cite{refactoringMiner, refactoringMiner1, refactoringMiner2} (F1: 97.96) and DesigniteJava 3.0~\cite{codeSmells} (F1: 98.82), which demonstrate high precision.

RQ2 assesses the impact of refactoring on code smells. However, our analysis operates at the level of the commit, which may contain other non-refactoring code changes. Therefore, the change in smell count ($\Delta$) is correlated with the refactoring event, but it is challenging to prove that the refactoring operation itself is the sole cause of the smell change. 
\section{Related Work}
\label{sec:related}
Recent years have seen the release of coding agents such as Claude Code~\cite{anthropic2025claudecode}, Copilot~\cite{microsoft2025copilot}, Cursor Agent~\cite{cursor2025docs}, Devin~\cite{devin2025ai}, and OpenAI Codex~\cite{openai2025codex}. Prior work has researched their use and effect on developer productivity, demonstrating that agents can help developers write code faster and reduce manual effort~\cite{peng2023the, wang2025ai, humaninloop}. However, while these studies assess general coding support, there is limited work on how agents perform refactoring in practice, the quality of changes introduced, or their impact on code maintainability.

Refactoring improves code readability, maintainability, and reduces technical debt~\cite{fowler2018refactoring, refactoring1}. Studies have researched the types and impact of developer refactorings in open source projects, including their effect on bug introduction, and long-term project health~\cite{bbagheri2022refactoring, nagy2022co, kim2012field, lacerda2020code}. These studies show how developers perform refactorings, but do not focus on agentic refactoring.

Prior work has developed tools for assisting and automating refactoring~\cite{fernandes2022liveref, kohler2019automated, maruyama2011security, ahmadi2022dqn}. However, no large-scale study compares agentic and developer refactoring in Java. We address this by analyzing refactorings across pull requests from 86 Java projects, providing a comparison of their behaviour and effects on code quality.

\section{Conclusion}
\label{sec:conclusion}
Our results show developers perform diverse refactorings focused on improving code structure and maintainability, while agents primarily perform annotation related edits. Agentic refactorings are often different from developer refactorings in terms of refactoring actions; however, their introduction of code smells across pull requests was statistically insignificant compared to humans, with the exception of Cursor. This indicates that agents are not introducing more code smells in refactoring commits than developers, despite performing significantly more refactoring.

Claude Code refactoring changes showed the largest skew toward annotation related refactorings, often introducing smells in the process. In contrast, Copilot demonstrated more balanced behaviour, performing fewer large-scale edits; however, it still makes more annotation related changes than developers. Cursor and Devin displayed limited refactoring activity, making it difficult to draw conclusions about their overall performance.

Overall, our findings suggest that while code agents can perform refactorings, their refactoring commits remain limited in their ability to improve code quality, introducing code smells at a similar rate as developers. These results highlight the need for future research into agentic refactoring that not only automates changes but also improves code quality through a more diverse and structural set of refactorings, rather than primarily focusing on annotations. For transparency and reproducibility, our Agentic and Human datasets, processed data, code, and supplementary material are available on Zenodo~\cite{datasetzenodo}.

\begin{acks}
This work was supported by the Natural Sciences and Engineering Research Council of Canada (NSERC).
\end{acks}

\bibliographystyle{ACM-Reference-Format}
\bibliography{bib}

\end{document}